\documentclass[journal = jctcce]{achemso}
\usepackage[version=3]{mhchem} % Formula subscripts using \ce{}

\title{Efficiency of energy funneling in the photosystem~II supercomplex of higher plants.}
\author{Christoph Kreisbeck}
\email{christophkreisbeck@gmail.com}
\affiliation{Department of Chemistry and Chemical Biology, Harvard University, 12 Oxford St, Cambridge, Massachusetts 02138, USA}
\author{Al\'an Aspuru-Guzik}
\affiliation{Department of Chemistry and Chemical Biology, Harvard University, 12 Oxford St, Cambridge, Massachusetts 02138, USA}
\email{aspuru@chemistry.harvard.edu}
\date{\today} 

\newcommand{\im}[0]{\mbox{i}}

\newcommand{\dt}[0]{\frac{\mbox{d}}{\mbox{d}t}}
\newcommand{\kb}[0]{k_{\rm B}}
\begin{document}
% \singlespacing

\begin{center}
\date{\today} 
\end{center}
%-----------------------------------------------------------------------------

\maketitle
% \begin{article}
\begin{abstract}
The investigation of energy transfer properties in photosynthetic multi-protein networks gives insight into their
underlying design principles.
Here, we discuss excitonic energy transfer mechanisms of the photosystem II (PS-II) C$_2$S$_2$M$_2$ supercomplex, which is the largest isolated functional unit of the photosynthetic apparatus of higher plants.
Despite the lack of a decisive energy gradient in C$_2$S$_2$M$_2$, we show that the energy transfer is directed by relaxation to low energy states. 
C$_2$S$_2$M$_2$ is not organized to form pathways with strict energetic downhill transfer, which has direct consequences on the transfer efficiency, transfer pathways and transfer limiting steps. 
The exciton dynamics is sensitive to small structural changes, which, for instance, are induced by the reorganization of vibrational coordinates. 
In order to incorporate the reorganization process in our numerical simulations, we go beyond rate equations and use the hierarchically coupled equation of motion approach (HEOM). 
While transfer from the peripherical antenna to the proteins in proximity to the reaction center occurs on a faster time scale, the final step of the energy transfer to the RC core is rather slow, and thus the limiting step in the transfer chain.
Our findings suggest that the structure of the PS-II supercomplex guarantees photoprotection rather than optimized efficiency. 
\end{abstract}

\section{Introduction}
Photosynthesis in which light is absorbed and converted in chemical energy is the most important process in nature. 
In higher plants the light-harvesting machinery is assembled of C$_2$S$_2$M$_2$ supercomplexes and networks of LHCII pigment-proteins \cite{dekker2005a,johnson2011a,kouvril2012a,duffy2013a} located in the grana membrane.
The C$_2$S$_2$M$_2$ supercomplex is formed by a dimeric photosystem II (PS-II) with moderately attached LHCII trimers and several minor complexes \cite{caffarri2009a}.
Energy transfer 
to the reaction center (RC) core pigments of PSII, in which the primary step of charge separation initializes an avalanche of photochemical reactions \cite{dekker2000a,diner2002a,kern2007a}, reaches remarkable efficiencies of up to $90\%$ \cite{caffarri2011a}.
However, it remains unclear of how such high efficiencies can be achieved in large and disordered systems.
In contrast to the photosynthetic apparatus of Green Sulfur Bacteria in which fast transfer is guaranteed by efficient energy funneling \cite{huh2014a}, microscopic derived Hamiltonians do not predict a decisive energy gradient among the individual proteins of the C$_2$S$_2$M$_2$ supercomplex \cite{novoderezhkin2011a,raszewski2008a,raszewski2008b,muh2012a,muh2014a}. 

Previous works describe the transfer kinetics with phenomenological models, and extract certain decay components such as the migration time (average time that it takes for an excitation to reach the RC) and trapping time by fitting to fluorescence decay lines \cite{broess2006a,broess2008a,croce2011a,chmeliov2014a}. 
Several rate limiting models are discussed in literature \cite{broess2008a,croce2011a,miloslavina2006a,tumino2008a}. Recent studies favor the so called transfer-to-trap limited kinetic model \cite{broess2008a,croce2011a,broess2006a, raszewski2008a} in which the transfer rate from the antenna complexes of PS-II to the RC is proposed to be the transfer limiting step.
However, different kinetic models can be fitted equally well to measured fluorescence decay curves \cite{bennett2013a}, and structure based models of energy transfer become necessary to shed light on the underlying transfer mechanisms. 
First microscopic simulations of the exciton dynamics in the C$_2$S$_2$M$_2$ supercomplex show that the overall transfer is driven by a complex interplay of multiple rates rather than through a single transfer-limiting step \cite{bennett2013a}. 

In pigment-protein complexes directionality of energy transfer is driven by energy relaxation.  
Variations in the energy bands of the individual proteins in C$_2$S$_2$M$_2$ are not as distinct as in other photosynthetic systems.
Nevertheless, the energy gradient in C$_2$S$_2$M$_2$ is not completely flat, and the pigments form a certain structure in the energetic layout.
For example CP43 and CP47 are lower in energy than the LHCII antenna complexes \cite{muh2012a, novoderezhkin2011a,raszewski2008a}. 
However, energy transfer in C$_2$S$_2$M$_2$ is not a cascade of downhill steps toward the reaction center.
Actually the pigments in the proximity of the RC core are the energetically lowest ones \cite{shibata2013a}. Therefore, the last transfer step to the trap needs to overcome an energy barrier which supports the proposed transfer-to-trap limited exciton dynamics in C$_2$S$_2$M$_2$.
The transfer limiting step to the RC core pigments, which is not anticipated in previous structure based simulations \cite{bennett2013a}, becomes more evident once we include the recently derived Hamiltonian of CP29 \cite{muh2014a}. The latter is substituted in Ref.~\cite{bennett2013a} by a LHCII monomer. We show that the minor complex CP29 modifies the pathway of energy flow and yields a relaxation channel which drives energy from the peripherical antenna towards pigments closer to PS-II.  

The transfer properties are sensitive to small structural modulations which is an immediate consequence induced by the flat energy gradient.
There are two major mechanisms which change the energetic structure: (i) static disorder in which site energies are subjected to random fluctuations on much slower time scales than the exciton dynamics and (ii) the reorganization process in which vibrational coordinates relax to a new equilibrium position after a vertical Franck-Condon transition to the excited state energy potential surface \cite{may2004a}. During this process the reorganization energy is dissipated in the protein environment. 
While the transfer times of an ensemble of individual disorder realizations are randomly distributed around some average value \cite{bennett2013a}, the reorganization process is a systematic effect pertaining to the dynamics in all realizations in the same way.

Due to the lack of the computational capability to carry out accurate calculations of the exciton dynamics, previous simulations of transfer time-scales in light-harvesting complexes (LHCs) employ a combined modified Redfield/generalized F\"orster rate equation approach \cite{renger2011a,yang2002a,novoderezhkin2012a,bennett2013a,zigmantas2006a}. However the combined modified Redfield/generalized F\"orster lacks the ability to simulate the reorganization process. In addition those models provide an \textit{ad} \textit{hoc} description of dynamic localization, and depend on an empirical cut-off parameter. Recently, a non-Markovian (ZOFE) quantum master equation description is employed to investigate robustness of transfer effiency and the importance of vibrational enhanced transfer in PS-II \cite{roden2015a}.
Here, we perform accurate simulations based on the hierarchically coupled equations of motion approach (HEOM) \cite{tanimura1989a,shi2009a,hu2011a, ishizaki2009c,tanimura2012a} which accurately incorporates the reorganization process and works for a wide parameter range for the coupling strength to the environment. 

Since the computational complexity of HEOM scales exponentially with increasing system size, 
novel algorithms based on optimized parallelization schemes have been developed \cite{kreisbeck2011a,struempfer2012a,kreisbeck2014a}.
The most efficient implementation \cite{kreisbeck2011a} employs the high compute throughput provided by modern graphics processing units (GPUs) for which a cloud computing version is hosted on nanohub.org \cite{kreisbeck2013a}.
GPU-HEOM is bound to the available GPU memory, and simulations are limited to intermediate sized systems. 
Here we overcome the memory limitation by using \textit{QMaster} \cite{kreisbeck2014a} which runs on various hardware architectures including GPUs and high memory multi core CPU architectures. 
We make use of the large CPU memory to benchmark the convergence of the hierarchy depth and use the high compute throughput of the GPUs for production runs. 

In Section~ \ref{sec:excitonmodel} we outline the structure of the Frenkel exction model for energy transfer in C$_2$S$_2$M$_2$. The technical aspects of the HEOM approach are stated in Section~\ref{sec:method}. 
After that, we continue with the discussion of time-scales of inter-protein transfer in the PS-II supercomplex (see Section~\ref{sec:discussion}). Finally, we investigate the impact of structural modifications on the transfer pathways and the transfer efficiency. 

\begin{figure}[t!]
\begin{center}
\includegraphics[width=0.4\textwidth]{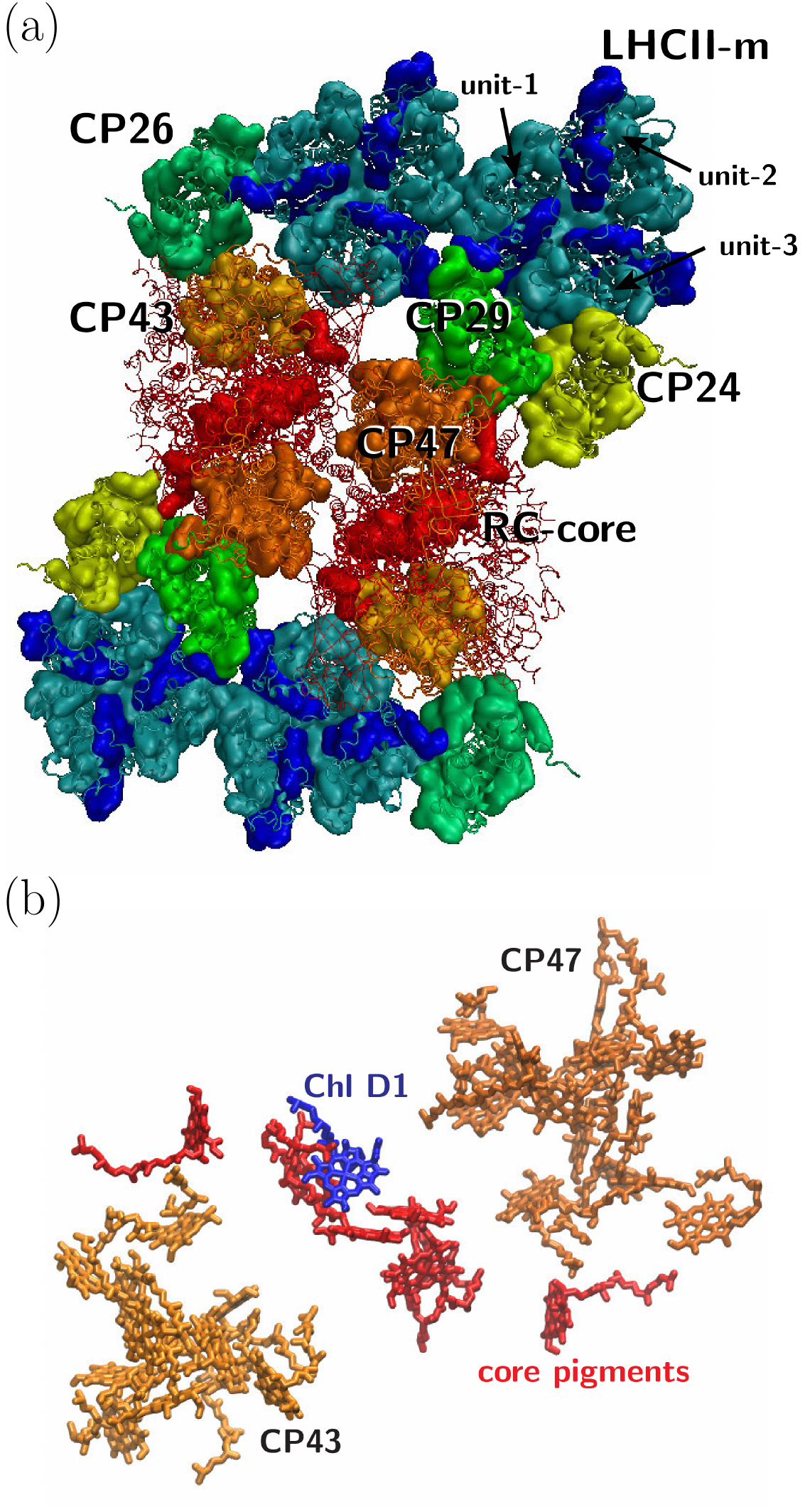}
\end{center}
\caption{\label{fig:PSIIsketch}\small (a) Sketch of the protein structure of the C$_2$S$_2$M$_2$ supercomplex.
The multiprotein complex contains 4 LHCII trimers, the minor complexes CP24, CP26 and CP29 which are connected to the PS-II \cite{caffarri2009a}.
(b) assembly of the pigments of PS-II composed of CP43, CP47 and the RC core. The primary step of charge separation is initiated through excitation of pigment Chl~D1 (see eq.~(\ref{chargeSep})).}
\end{figure}  

\section{Exciton Model}\label{sec:excitonmodel}
The orientation of the individual proteins of the C$_2$S$_2$M$_2$ supercomplex is determined by a projection map at 12 \AA{} resolution \cite{caffarri2009a}. 
The C$_2$S$_2$M$_2$ supercomplex, which structure is depicted in Fig.~\ref{fig:PSIIsketch}(a), comprises four LHCII trimers, six minor light-harvesting complexes and a dimeric PS-II core complex. 
Absorbed light in the outer LHCII antenna complexes is transfered via the minor complexes CP24, CP26, and CP29 to CP47 and CP43 of PS-II. 
The transfer process is completed by irreversible charge separation triggered in the RC core.  
Electron transfer in the RC core is described phenomenologically by radical pair states RP1 , RP2 and RP3 \cite{holzwarth2006a,raszewski2008a}. We assume that the primary step of charge separation is initiated through the electronically excited core pigment Chl$_{\rm D1}$ (the location of Chl$_{\rm D1}$ in PS-II is illustrated in Fig.~\ref{fig:PSIIsketch}(b)) and described by the rate equation
\begin{equation}\label{chargeSep}
 \mbox{Chl}^\ast_{\rm D1}\mbox{Pheo}_{\rm D1}\stackrel{\Gamma_{\rm RP1}}{\longrightarrow}\mbox{Chl}^+_{\rm D1}\mbox{Pheo}_{\rm D1}^-.
\end{equation} 
We neglect backward rates since fluorescence decay lines suggest that charge recombination occurs on a much slower time scale than primary electron transfer \cite{raszewski2008a}. 
Within this limit we model primary charge separation as irreversible exciton trapping. 
In literature also more sophisticated models are discussed which include multiple pathways of charge separation \cite{novoderezhkin2011b,romero2014}.

We describe energy transfer in the C$_2$S$_2$M$_2$ supercomplex within a Frenkel exciton Hamiltonian for which we assume that only one of the pigments is excited at once. The Hamiltonian of the single exciton manifold reads
\begin{equation}\label{eq:Hex}
H_{\rm ex}=\sum_{m=1}^N \epsilon_m^0 |m\rangle\langle m| + \sum_{m>n} J_{mn}(|m\rangle\langle n|+|n\rangle\langle m|).
\end{equation}
Here $|m\rangle$ denotes the state in which pigment~$m$ is excited while the other pigments remain in the electronic ground state. 
For the inter-site couplings $J_{mn}$ we distinguish between intra-complex and inter-complex coupling terms depending whether or not pigments $m$ and $n$ are located within the same protein.  
We use the same parameter for the exciton Hamiltonian which is constructed by Bennett $et$ $al.$ in Ref.~\cite{bennett2013a}.
Recently the Hamiltonian for CP29 has been resolved \cite{muh2014a} which is in Ref.~\cite{bennett2013a} replaced by a LCHII monomer (without pigment Chl 605).
In order to isolate of how much the new CP29 Hamiltonian influences transfer and to compare the HEOM results with previous approximate modified Redfield/generalized F\"orster simulations \cite{bennett2013a}, we carry out calculations for both models: (i) with the CP29 Hamiltonian and (ii) with the LHCII monomer substitution. 

The pigments are coupled to the protein environment modeled by a set of independent harmonic oscillators 
\begin{equation}
\mathcal{H}_{\rm phon}=\sum_{m,i}\hbar\omega_i b_{i,m}^\dag b_{i,m},
\end{equation}
and we assume a linear coupling of the exciton system to the vibrations 
\begin{equation}
\mathcal{H}_{\rm ex-phon}=\sum_m |m\rangle\langle m|\,\sum_i\hbar\omega_{i,m}d_{i,m}(b_{i,m}+b_{i,m}^\dag).
\end{equation}
The reorganization energy $\mathcal{H}_{\rm reorg}=\sum_m\lambda_m |m\rangle\langle m|$, with 
$\lambda_m=\sum_i\hbar\omega_{i,m}d_{i,m}^2/2$ is added to the exciton Hamiltonian eq.~(\ref{eq:Hex}). 
We define the site energies as $\varepsilon_m=\varepsilon_m^0+\lambda_m$. 
The phonon mode dependent coupling strength is captured by the spectral density
 \begin{equation}\label{eq:SpecDens}
 J_m(\omega)=\pi\sum_\xi \hbar^2\omega_{\xi,m}^2 d_{\xi,m}^2\delta(\omega-\omega_{\xi,m}).
\end{equation}
Frequently, the reorganization energy and the spectral density are assumed to be site independent. However for the C$_2$S$_2$M$_2$ supercomplex 
each of individual protein has its own reorganization energies and own form for the spectral density  \cite{bennett2013a}. 
The spectral density for LHCII is extracted from fluorescence line narrowing spectra. Since the experimental spectra cannot distinguish between the Chl$a$ and Chl$b$ pigments we assume for both the same spectral density composed of 48 vibrational peaks \cite{novoderezhkin2004a,novoderezhkin2010a}.
Transfer times are not much affected by the structures in the spectral density and a coarse grained Drude-Lorentz spectral density is appropriate to describe energy transfer in LHCII \cite{kreisbeck2014a}.
Microscopic details for the spectral densities of the minor complexes and PS-II are not know. The structure of CP29 is similar to the one of a LHCII monomer. Thus we assume that the spectral density of CP29 can be substituted with the LHCII spectral density\cite{bennett2013a, muh2014a}. 
For CP47 and the RC core pigments $\lambda=38.64$~cm$^{-1}$ and $\lambda=50.23$~cm$^{-1}$, respectively are suggested as reasonable values for the reorganization energy \cite{raszewski2008a}. For the RC core pigments also a higher reorganization energy is discussed \cite{novoderezhkin2011b,novoderezhkin2005a}.
The explicit form and parameter for the spectral densities used in this manuscript are listed in Appendix \ref{ParamSpecDens}.

\begin{figure*}[t!]
\begin{center}
\includegraphics[width=0.9\textwidth]{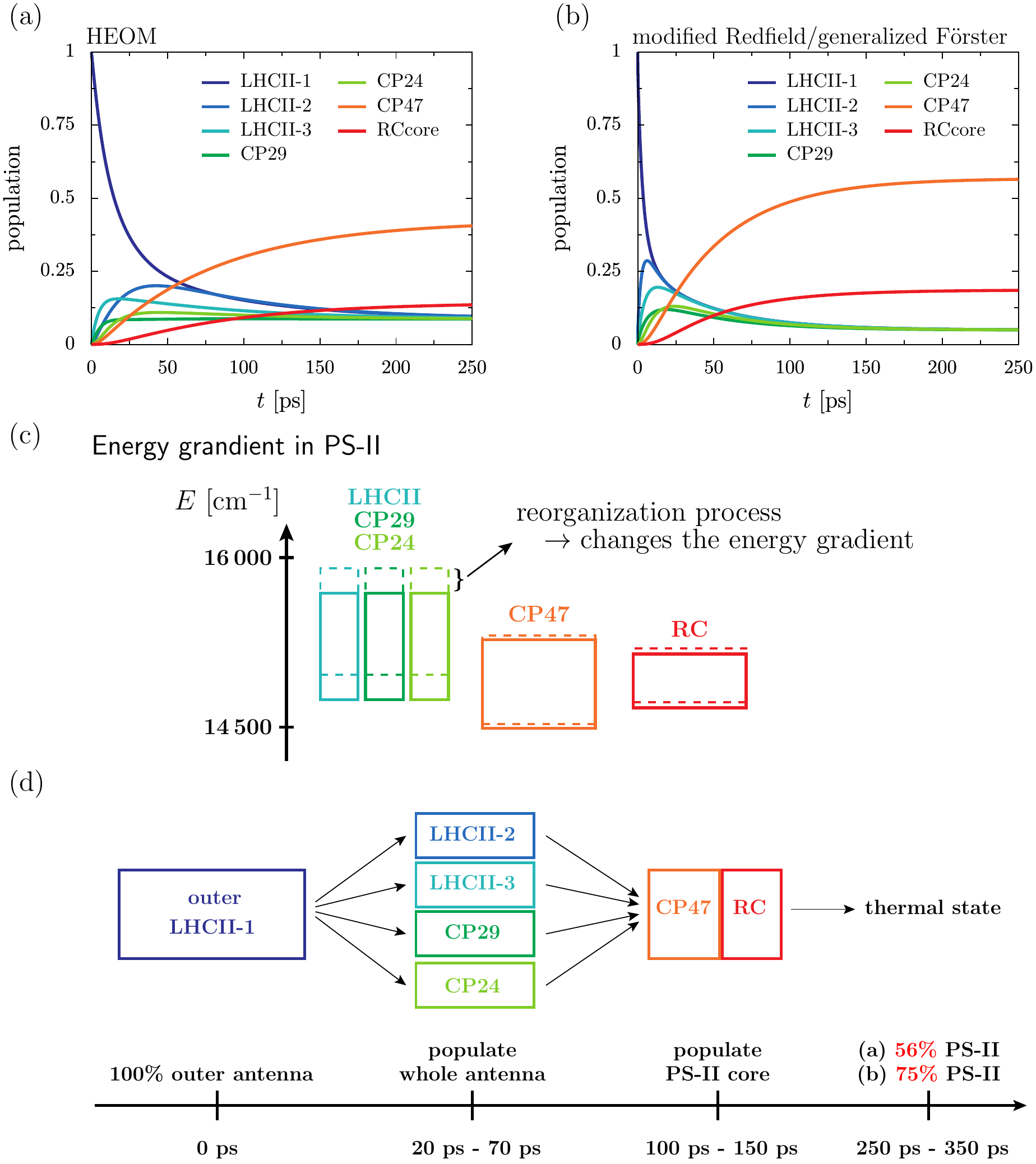}
\end{center}
\caption{\label{fig:Plot_Module_Populations_Average}\small  Aggregated  populations at $T=277$~K in absence of trapping in the 93 site network comprising LHCII-m, CP24, CP29, CP47 and the RC-core. The initial state is given by the highest exciton state within the domain of $\mathcal{H}_{\rm strong}$ which dominantly excites pigment Chl$b$~606 of the LHCII-m unit-1. Depicted are the population dynamics within the (a) HEOM and (b) combined modified Redfield/generalized F\"orster approach. (c) Sketch of the layout of the exciton energy bands, with (dashed boxes) and without (solid boxes) the reorganization energy. (d) Illustrates rough estimates for the time scales of how energy spreads across the individual proteins.   
}
\end{figure*}

We include the primary step of charge separation phenomenologically as irreversible population trapping, which we incorporate by anti-Hermitian parts in the Hamiltonian
\begin{equation}
 \mathcal{H}_{\rm trap} = -i\hbar\Gamma_{\rm RP1}/2\,|\mbox{Chl}_{\rm D1}\rangle\langle\mbox{Chl}_{\rm D1}|,
\end{equation}
where $\Gamma_{\rm RP1}$ defines the rate of the primary charge separation.
In a similar way we incorporate exciton losses 
\begin{equation}
 \mathcal{H}_{\rm loss} = -i\hbar\Gamma_{\rm loss}/2\,\sum_m|m\rangle\langle m|.
\end{equation}
where we assume exciton lifetimes of $(\Gamma_{\rm loss})^{-1}=2$~ns. 
We characterize transfer properties by the transfer efficiency 
\begin{equation}
 \eta=\int_0^{t_{max}}\mbox{d}t\, \Gamma_{\rm RP1}  \langle\mbox{Chl}_{\rm D1}|\rho(t)|\mbox{Chl}_{\rm D1}\rangle\
\end{equation}
and average transfer time
\begin{equation}
 \langle t \rangle=\Gamma_{\rm RP1}/\eta \int_0^{t_{\rm max}} \mbox{d}t\, t \langle\mbox{Chl}_{\rm D1}|\rho(t)|\mbox{Chl}_{\rm D1}\rangle.
\end{equation}
For numerical evaluations we replace the upper integration limit by $t_{max}$ which is chosen such that the total population within the pigments of the C$_2$S$_2$M$_2$ supercomplex has dropped below $0.001$. 

\section{Method}\label{sec:method}
We evaluate the exciton dynamics within the hierarchically coupled equation of motion (HEOM) method \cite{tanimura1989a,ishizaki2009c,tanimura2012a}. HEOM is an open quantum system approach which treats the coupling to the vibrational modes as a bath. The time evolution of the total density operator $R(t)$, which characterizes the degrees of freedom of the exciton system as well as the ones of the phonon bath, is governed by the Liouville equation
\begin{equation}\label{eq:Liouville}
 \dt {R}(t)= - \frac{\im}{\hbar}[{\mathcal{H}}(t),{R}(t)]=- \frac{\im}{\hbar}{\mathcal{L}}(t){R}(t).
\end{equation}
We assume that the total density operator factorizes at initial time $t_0=0$ in system
and vibrational degrees of freedom $R(t_0)=\rho(t_0)\otimes\rho_{\rm phon}(t_0)$. 
To get the time evolution of the reduced density matrix $\rho(t_0)$,   
we trace out the bath degrees of freedom
\begin{equation}\label{HEOM2}
 {\rho}(t)=\langle\mbox{T}_+\,\exp\Big(-\frac{\im}{\hbar}\int_0^t \mbox{d}s \,{\mathcal{L}}(s) \Big)\rangle{\rho}(0) .
\end{equation}
By employing second order cumulant expansion, using a Drude-Lorentz spectral density $J_m(\omega)=2\lambda_m \frac{\omega \gamma_m}{\omega^2+\gamma_m^2}$ in combination with a high temperature approximation $\hbar\gamma_m/\kb T<1$, we cast the time non-local eq.~(\ref{HEOM2}) into a hierarchy of coupled time local equations of motion
\begin{eqnarray}\label{HEOM12}
 \frac{\mbox{d}}{\mbox{d} t}\sigma^{(n_1, ...,n_N)}(t)&=&
\big(-\frac{\im}{\hbar}\mathcal{L}_{\rm ex}
-\sum_m n_m\gamma
\big)\sigma^{(n_1, ...,n_N)}(t)\nonumber\\
&&+\sum_m
\frac{\im}{\hbar} V_m^{\times}
\sigma^{(n_1, ...,n_m+1,...,n_N)}(t)\nonumber \\
&&+\sum_mn_m\theta_m\sigma^{(n_1, ...,n_m-1,...,n_N)}(t).
\end{eqnarray}
where we define $\rho(t)=\sigma^{\vec{0}}(t)$, $\theta_m=\im\Big(\frac{2\lambda}{k_B T\hbar}{V}_{m}^{\times}(t)
-\im\lambda\gamma {V}_{m}^{\circ}(t)\Big)$, $V_m^\times\,\bullet = [V_m,\,\bullet]$, $V_m^\circ\,\bullet = \{V_m,\,\bullet\}$ and ${V}_{m}=|m\rangle\langle m|$. The hierarchy can be truncated for a sufficiently large hierarchy level $\sum_m^{N_{\rm sites}}n_m>N_{\rm max}$. Convergence of the hierarchy can be tested by comparing deviations in the dynamics with increasing truncation level.
To increase the accuracy of the high temperature approximation of HEOM \cite{olsina2013a} we include additional correction terms \cite{ishizaki2009a} for which we replace 
\begin{eqnarray}\label{HEOM16}
 {\cal L}_{\rm ex}&\rightarrow&\ {\cal L}_{\rm ex}-\sum_{m=1}^{N}\frac{2 \lambda}{\beta\hbar^2}\frac{2
 \nu}{\gamma_1^2-\nu^2}V_m^\times V_m^\times \nonumber \\
 {\Theta}_{m}&\rightarrow&\ {\Theta}_{m}-\frac{2 \lambda}{\beta\hbar}\frac{2
 \nu^2}{\gamma_1^2-\nu^2}V_m^\times.
\end{eqnarray}
For structured spectral densities a similar hierarchical expansion has been derived which relies on a decomposition of the spectral density in terms of shifted Drude-Lorentz peaks \cite{kreisbeck2012a,kreisbeck2013b} or underdamped Brownian oscillators \cite{tanimura2012a}.

\section{Discussion}\label{sec:discussion}
Together with LHCII complexes, the C$_2$S$_2$M$_2$ supercomplex aggregates as a large photosynthetic network in the grana membrane. 
For each C$_2$S$_2$M$_2$ supercomplex there are about six additionally loosely bound LHCII trimers \cite{duffy2013a}, which form a large antenna system with densely packed chlorophylls. 
Energy is either absorbed in the pool of loosely bound LHCII trimers and then transfered to one of the peripherical LHCIIs of the C$_2$S$_2$M$_2$ supercomplex or absorbed directly in the LHCII trimers of C$_2$S$_2$M$_2$. 
Further, to some extend energy is absorbed in the minor complexes and PS-II. 
We expect that the contribution of light absorption in the minor complexes and PS-II to the photosynthetic yield is of less importance, since most of the photoactive area in the grana membrane is covered by the LHCIIs.
Thus, to reach a high photosynthetic yield fast and efficient transfer from the LHCIIs toward the RC core pigments of PS-II becomes indispensable. 

In the following we investigate average transfer times and the efficiency of energy transfer from the peripherical LHCII-m monomeric unit labeled as unit-1 in Fig.~\ref{fig:PSIIsketch} to the reaction center in which the primary step of charge separation takes place. 
We employ the presence of a certain amount of symmetry along x-axis and y-axis and reduce the system to a multi-protein network composed of LHCII-m, CP24, CP29, CP47 and the RC-core, comprising 93 pigments in total.

\subsection{Energy gradient drives directionality}
First, we keep track of the population dynamics in absence of trapping and energy losses. We highlight of how energy spreads among the different protein complexes which, as we will analyze in detail, is driven by energy gradients in the pigments of the C$_2$S$_2$M$_2$ supercomplex. Further, we explore the influence of the reorganization process on the exciton dynamics. 

Following, we investigate the deficiency of the approximate combined modified Redfield/gener\-alized F\"orster method by comparing the population dynamics obtained from the combined method with the HEOM results. 
The combined modified Redfield/generalized F\"orster approach divides the exciton Hamiltonian eq.~(\ref{eq:Hex}) into a strongly coupled part $\mathcal{H}_{\rm strong}$ (associated with strongly coupled domains) and a weakly coupled part. Hereby $\mathcal{H}_{\rm strong}$ is defined as the exciton Hamiltonian in which the inter-site couplings $J_{nm}$ are set to zero if the coupling strength drops below a certain threshold value. We follow Ref.~\cite{bennett2013a} and use a threshold of 15~cm$^{-1}$. The intra domain dynamics is then modeled by modified Redfield, while the inter-domain transfer is described by generalized F\"orster theory. 
Since the choice of initial conditions of the combined method is restricted to eigenstates of certain domains in $\mathcal{H}_{\rm strong}$ we set the highest energy state of the domain which predominantly populates pigment Chl$b$~606 of the LHCII-m unit-1 as initial condition. To allow for comparison, we use the same initial condition for the HEOM calculations.

Figure~\ref{fig:Plot_Module_Populations_Average}(a) depicts the aggregated population at the individual protein complexes obtained within HEOM. Convergence of the hierarchy depth is verified by comparing with a higher truncation level (see Appendix~\ref{ConvergHEOM}). 
Overall energy transfer and directionality is driven by energy relaxation along energy gradients within the C$_2$S$_2$M$_2$ supercomplex. LHCII and the minor complexes (modeled by LHCII monomers without Chl~605) exhibit the highest energies while CP47 and the RC core pigments are lower in energy. 
The exciton, initially located at the unit-1 LHCII-m monomer, spreads over the complete LHCII-m trimer and populates the minor complexes. 
The fast initial spread, shows as maxima in the aggregated populations at LHCII-m units-2 and 3.
The highest population at the unit-3 is obtained after about 18~ps while the maximum population at unit-2 is reached a bit later at about 43~ps. 
The high population of the LHCII-m unit-2 indicates that energy transfer does not exclusively proceed along pathways corresponding to the shortest distance to PS-II and the RC. 
The minor complexes are populated on a similar time-scale as the monomeric LHCII-m units.
On a slower timescale CP47 and the RC core of PS-II get populated, and finally after about 250~ps \-- 350~ps the system reaches steady state in which energy relaxation drives the population to the energetically low exciton states at CP47. A schematic sketch summarizing the rough estimates for the energy transfer time-scales is given in Fig.~\ref{fig:Plot_Module_Populations_Average}(d).

The dynamics of the combined modified Redfield/generalized F\"orster approach (Fig.~\ref{fig:Plot_Module_Populations_Average}(b)) diverges from the HEOM results in several aspects. Overall relaxation is  overestimated by the combined modified Redfield/generalized F\"orster approach. Especially transfer to LHCII-m unit-2 is about seven times faster and already at 6~ps there is 0.29 population at the unit-2. Further, unit-2 gets populated ahead of unit-3. Therefore the pathway of how energy spreads over the monomeric units of the LHCII-m trimer is reversed when compared to the HEOM calculation and thus the 
combined method does not predict reliable pathways of energy flow during the first picoseconds.
However, the main difference is in the stationary population which is not only approached faster (at about 150~ps \-- 250~ps) but predicts a much higher aggregated population at CP47 and the RC. The combined modified Redfield/generalized F\"orster approach overestimates the efficiency of energy funneling towards the PS-II. 

\begin{figure}[t!]
\begin{center}
\includegraphics[width=0.45\textwidth]{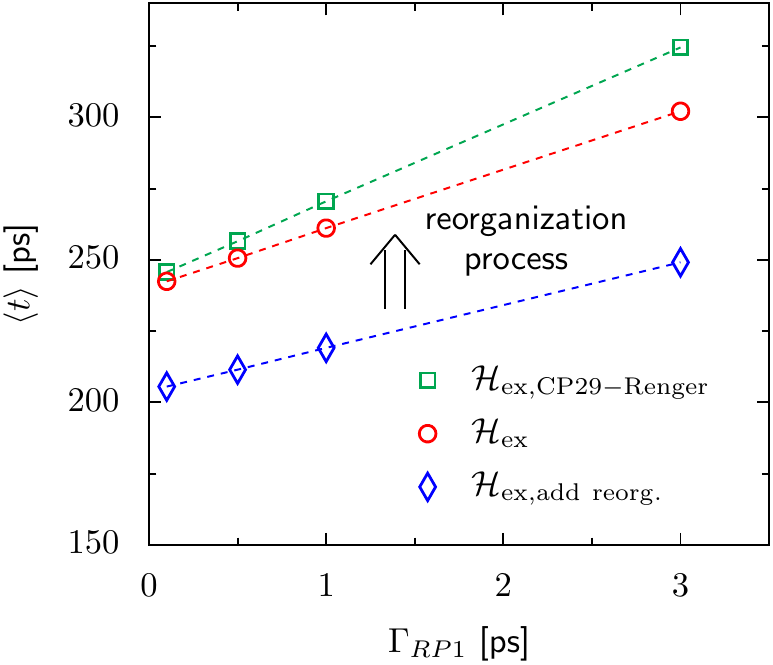}
\end{center}
\caption{\label{fig:Plot_averageTrapping}\small Trapping time evaluated for various rate constant of primary charge separation $\Gamma_{\rm RP1}$ at T=277~K. The transfer time is given as average over different initial conditions corresponding to eigenstates of the isolated LHCII-m unit-1 monomer. We investigate changes in the transfer time induced by structural changes in the Hamiltonian. We compare three different scenarios, (i) $\mathcal{H}_{\rm ex, CP29-Renger}$ for which we use the CP29 Hamiltonian of Renger $et$ $al.$ Ref.~\cite{muh2014a}, (ii) $\mathcal{H}_{\rm ex}$ for which the CP29 is substituted by a LHCII monomer (without Chl~605) and (iii) $\mathcal{H}_{\rm ex, add\ reorg.}$ for which we add the reorganization energy to the site energies of LHCII-m and the minor complexes of $\mathcal{H}_{\rm ex}$}
\end{figure} 

\begin{figure*}[t]
\begin{center}
\includegraphics[width=0.8\textwidth]{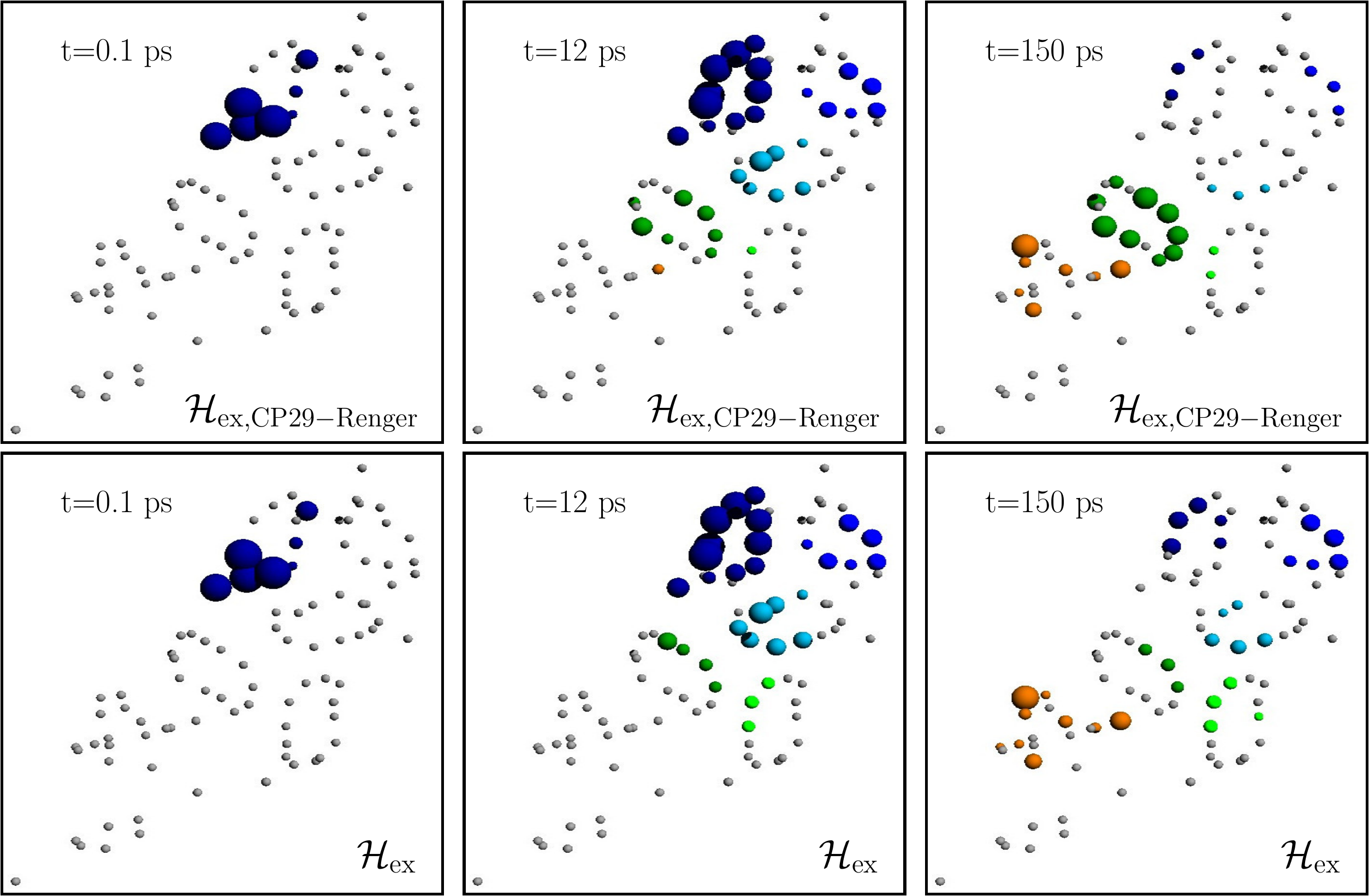}
\end{center}
\caption{\label{fig:Plot_snapshots}\small Snapshots of the exciton dynamics in presence of population trapping in the reaction center ( $\Gamma_{\rm RP1}^{-1}$=0.5~ps) at $T=277$~K. The spheres represent the position of the individual pigments while the radii reflect the population at each pigment (we employ a arctan scale). The color encodes the population at the individual protein complexes. Spheres in gray indicate pigments with less than 0.0079 population. As initial condition we use the highest eigenstate of the isolated LHCII-m unit-1.
The upper and lower panels show results for two different Hamiltonians, $\mathcal{H}_{\rm ex, CP29-Renger}$ and $\mathcal{H}_{\rm ex}$ respectively. Both differ in the structure of the minor complex CP29.}
\end{figure*} 

To understand the discrepancy in the stationary population we need to investigate the energetic layout of the C$_2$S$_2$M$_2$ supercomplex.
The stationary state follows typically a thermal Boltzmann distribution. 
However, the situation becomes more complicated in presence of the reorganization process in which the reorganization energy dissipates during the dynamics which modifies the energetic layout and affects the thermal population. 
The boxes in Fig.~\ref{fig:Plot_Module_Populations_Average}(c) indicate the extension of the exciton bands of the isolated proteins. 
The dashed lines correspond to the situation where the site energies comprise of the bare excitation energy plus the reorganization energy. 
Due to the reorganization process the energetic structure changes during the dynamics and the energy of the proteins is lowered by the reorganization energy. Especially the band of the monomeric LHCII-m units and the band of the minor complexes shifts to lower energies while the small reorganization energies at CP47 and RC induce only minor modifications. In total the already flat energy gradient gets even more flattened. 
This has a significant impact on the thermal population. 
Without the reorganization process (dashed lines) we expect a thermal population of about 0.75 at the pigments of PS-II. Taking into account the reorganization process (solid line) reduces the efficiency of energy funneling and only a population of 0.56 accumulates at PS-II. 
Our analysis is in consistency with the findings for the population dynamics and explains the strong deviations in the stationary state between HEOM and the combined modified Redfield/generalized F\"orster method. We note that for the combined modified Redfield/generalized F\"orster method the effect of reorganization energy on the thermal population can be corrected by subtracting the reorganization energy from the exciton Hamiltonian prior to the dynamics. This is based on the assumption that the reorganization energy dissipates on an infinitely fast time-scale.

\subsection{Structural variations modify the transfer efficiency}
In the following we investigate how minor structural modifications in the C$_2$S$_2$M$_2$ supercomplex influence transfer properties such as transfer efficiency and average transfer time.
As we have discussed in detail in the previous section, one mechanism that induces structural changes is the reorganization process. 
Here, we continue the discussion and examine how much the reorganization process affects transfer efficiency. 
Another aspect is the influence of the replacement of the minor complexes with LHCII monomeric units on the transfer properties.  
For instance the recently derived exciton Hamiltonian of CP29 shows various differences from the exciton system of a LHCII monomer \cite{muh2014a}. 

We incorporate the primary step of charge separation by irreversible energy trapping as is described in section \ref{sec:excitonmodel}. 
Different values for the rate constant of primary charge separation $\Gamma_{\rm RP1}$ have been predicted from fits to fluorescence decay lines, ranging from $\Gamma_{\rm RP1}^{-1}$=0.1~ps \cite{raszewski2008a} to $\Gamma_{\rm RP1}^{-1}$=0.64~ps \cite{bennett2013a}.
Pump-probe spectra predict even larger time constants for the Pheo reduction of about 3~ps \cite{holzwarth2006a}. 
We do not explicitly take into account mechanisms of photoprotection and quenching and phenomenologically describe exciton losses by assuming an exciton lifetime of $(\Gamma_{\rm loss})^{-1}=2$~ns.

In the following we carry out HEOM simulations in which we include trapping and energy losses. 
To investigate the effects of the reorganization process on the energy transfer times, we slightly modify the Hamiltonian of the C$_2$S$_2$M$_2$ supercomplex in a benchmark calculation for which we artificially restore the original energy gradient across the pigment proteins by adding the reorganization energy of 220~cm$^{-1}$ to the site energies of LHCII and the minor complexes. We neglect the minor energetic changes induced by the reorganization process at the pigments of CP47 and the RC and denote the modified Hamiltonian as $\mathcal{H}_{\rm ex, add\ reorg.}$. 
Relaxation time scales in the population dynamics are hardly affected by the shifts in the site-energies, but the thermal state adjusts now according to the modified energy gradient. 
For $\mathcal{H}_{\rm ex, add\ reorg.}$ we obtain a similar thermal state in the population dynamics with high population at the PS-II pigments (0.81) as is predicted by the calculations with the combined modified Redfield/generalized F\"orster method.  
The small deviations largely result from the fact that we did not add additional reorganization energies to the site energies of CP47 and RC. 

The transfer time as function of trapping rate follows a linear trend for the considered parameter regime as is illustrated in Fig.~\ref{fig:Plot_averageTrapping}.
We assume that initially the exciton is located at the LHCII-m unit-1, and we populate the initial density matrix according to eigenstates of the isolated LHCII monomeric unit. The shown results correspond to transfer times averaged over all 14 exciton states used as initial condition.
Transfer times (efficiency) for the C$_2$S$_2$M$_2$ supercomplex (marked by the red circles) are in the range between 242~ps (88.0\%) and 302~ps (85.2\%), depending on the  trapping rate $\Gamma_{\rm RP1}$. $\mathcal{H}_{\rm ex, add\ reorg.}$ exhibits a more efficient energy energy funneling towards the pigments at PS-II and therefore transfer is faster by about 36~ps to 53~ps. 
Previous calculations based on the combined modified Redfield/generalized F\"orster method predict transfer times of about 200~ps for transfer from peripherical domains to the Chl~D1 in the RC \cite{bennett2013a}. This is in good agreement with our results for $\mathcal{H}_{\rm ex, add\ reorg.}$ which yields a transfer time of 211~ps for trapping rate of $\Gamma_{\rm RP1}=0.5$~ps which is similar to the one used in Ref.~\cite{bennett2013a}.

\begin{figure}[t]
\begin{center}
\includegraphics[width=0.45\textwidth]{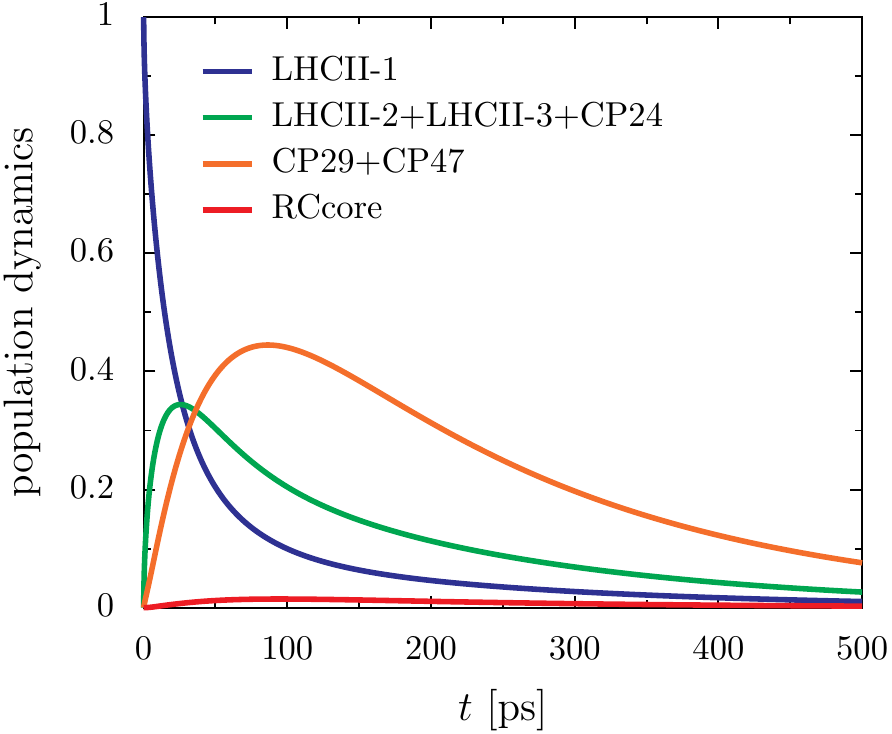}
\end{center}
\caption{\label{fig:Plot_Module_Populations_Average_eff}\small Aggregated populations for $\mathcal{H}_{\rm ex, CP29-Renger}$ in presence of trapping ($\Gamma_{\rm RP1}=0.5$~ps) at 277~K. Energy accumulates at low-energy bottleneck states at CP29 and CP47 limiting transfer to the RC.}
\end{figure}

\subsection{Minor complex CP29 guides transfer}
Since already small structural modifications such as the reorganization process alters the energy transfer efficiency, we expect that the substitution of the minor complexes by LHCII monomers may also significantly affect the energy transfer properties. In this section we use the Hamiltonian of CP29 derived by Renger $et$ $al.$ Ref.~\cite{muh2014a} instead of the LHCII monomer replacement. 
We denote the new Hamiltonian as $\mathcal{H}_{\rm ex, CP29-Renger}$, while the previous situation with the LHCII monomer substitution is referred to as $\mathcal{H}_{\rm ex}$. 

For $\mathcal{H}_{\rm ex, CP29-Renger}$ the pigments of CP29 form the lowest energy band in the energetic layout of the C$_2$S$_2$M$_2$ supercomplex. 
This has a two-fold implication on the transfer process. Firstly, the energy gradient between the outer LHCII antenna and the minor complex CP29
gives rise to an additional grade of directionality and 
supports fast transfer from the peripherical LHCII-m trimer to CP29.
The minor complex CP29 presumably acts as exit marker which guides energy from the outer antenna towards pigments closer to the reaction center. 
Secondly, the pigments of CP29 and CP47 form a spatially extended region of low-energy states and hence energy accumulates at pigments in proximity to the RC, while the final transfer step to the RC core pigments is energetically uphill and therefore slow. 
Overall the two effects result in a slightly slower energy transfer within the C$_2$S$_2$M$_2$ supercomplex while including the CP29 Hamiltonian, see Fig.~\ref{fig:Plot_averageTrapping}. 
For large trapping rates $\Gamma_{\rm RP1}>1$~ps the slow down of the energy transfer gets more pronounced. 

Figure \ref{fig:Plot_snapshots} charts snapshots of the exciton dynamics. The upper (lower) panels correspond to $\mathcal{H}_{\rm ex, CP29-Renger}$ ($\mathcal{H}_{\rm ex}$).  The radius of the colored spheres represents the population at each pigment. For better visualization we use an arctan scale. The spheres are uncolored if the population remains below 0.0079. Initially the highest eigenstate of the LHCII-m unit-1 is excited.  
Both Hamiltonians show a fast spread of the energy and at 12~ps the energy distributes across the whole LHCII-m timer. 
While $\mathcal{H}_{\rm ex}$ distributes population equally among the minor complexes $\mathcal{H}_{\rm ex, CP29-Renger}$ yields a more directed energy transfer towards the CP29 and CP47. 
For longer times of 150~ps energy accumulates at the low energy states at CP29 and CP47 for $\mathcal{H}_{\rm ex, CP29-Renger}$ and thus forms a bottleneck for transfer to the RC.  The bottleneck is less pronounced for $\mathcal{H}_{\rm ex}$.
The RC pigments do not show significant population at any time since as soon as energy enters the RC there is fast transfer to Chl~D1 and the fast time-scale of primary charge separation leads to the trapping of the population.

The rate limiting step in the transfer chain is the energetically up-hill transfer to the RC core. This is illustrated best in the aggregated population dynamics in presence of trapping in the reaction center, Fig.~\ref{fig:Plot_Module_Populations_Average_eff}. 
We obtain a fast decay of population in the LHCII-m and after 100~ps more than 0.75 of the population has left the LHCII-m trimer. At the same time about 0.44 of population accumulates in CP29 and CP47. After 300~ps still 0.2 of the population remains at CP29 and CP47.
 
\section{Conclusion}

With \textit{QMaster}, a high-performance implementation of the HEOM method, accurate calculations of excitonic energy transfer in multi-protein photosynthetic functional units such as the C$_2$S$_2$M$_2$ supercomplex become feasible. 
We investigate transfer times and transfer efficiency of energy conversion within the primary step of charge separation. 

The general concept behind energy transfer in C$_2$S$_2$M$_2$ is given by energy relaxation. 
Due to the flat energy gradient across the proteins,  small structural changes such as the reorganization of the molecular coordinates within the excited potential energy surface, affect the energy transfer process. 
The impact of the reorganization process is rather significant and energy relaxation drives much less population to CP47 and the RC than expected from the site energies of the Hamiltonian.  
The reorganization process induces a noticeable drop in the transfer efficiency of about 1.8\% to 2.6\% in absolute numbers for a 2~ps exciton lifetime, and thus cannot be neglected in simulations of energy transfer in large multi-protein complexes. 

Our simulations suggest that the minor complex CP29 acts as exit marker and adds directionality to the energy transfer from the peripherical LHCII to the proteins in the proximity to the RC core. 
The C$_2$S$_2$M$_2$ supercomplex is not optimized for efficient transfer. Energy accumulates in low energy states at CP29 and CP47, while the final transfer step needs to overcome an energy barrier and therefore is slow. 
Thus the energy transfer exhibits the character of a transfer-to trap limited model. 
In conclusion, within our model, we show that the structural layout of C$_2$S$_2$M$_2$ is not optimized for efficient transfer and 
suggests that photoprotection considerations are very relevant. The extension of accurate HEOM models to this case is possible and a promising direction for future research.

\begin{acknowledgement}
The authors would like to thank Dr. Kapil Amernath and Dr. Doran Bennett for helpful discussion. We thank Dr. Doran Bennett for providing the atomistic structure and the Hamiltonian of C$_2$S$_2$M$_2$.
This work was supported as part of the Center for Excitonics, an Energy Frontier Research Center funded by the U. S. Department of Energy, Office of Science, Office of Basic Energy Sciences under Award Number DE-SC0001088.
We thank Nvidia for support via the Harvard CUDA Center of Excellence. The computations in this paper were run on the Odyssey cluster supported by the FAS Division of Science, Research Computing Group at Harvard University.
\end{acknowledgement}

\begin{suppinfo}
\appendix
\section{Parameter of the spectral densities}\label{ParamSpecDens}
The coupling of each pigments to the vibrational environment is described by a Drude-Lorentz spectral density 
\begin{equation}
 J_m(\omega)=2\lambda_m \frac{\omega \gamma_m}{\omega^2+\gamma_m^2}.
\end{equation}
The paramter $\lambda_m$ and $\gamma_m$ for the individual pigments are listed in Table~\ref{tab:SpectralDensity}.
\begin{table}[h]
 \begin{center}
\begin{tabular}{c|c|c} 
pigment protein & $\lambda$ & $\gamma^{-1}$ \\
\hline
 LHCII, CP29 and CP24 & 220~cm$^{-1}$ & 15~fs \\
 CP47 & 38.64~cm$^{-1}$ & 50~fs \\
 RC core & 50.23~cm$^{-1}$ & 50~fs \\
 \caption{\label{tab:SpectralDensity}Parameter for the spectral density}
\end{tabular}
\end{center}
\end{table}

\section{Convergence of HEOM}\label{ConvergHEOM}
In order to test convergence of HEOM the dynamics is compared for different truncation levels as is charted in Fig.~\ref{fig:Plot_Module_TestConvergence}. Initially the highest exciton state of LHCII-m unit~1 monomer is excited. A truncation level of $N_{\rm max}=2$ yields sufficient accuracy for the relaxation time-scales.
\begin{figure}[t]
\begin{center}
\includegraphics[width=0.5\textwidth]{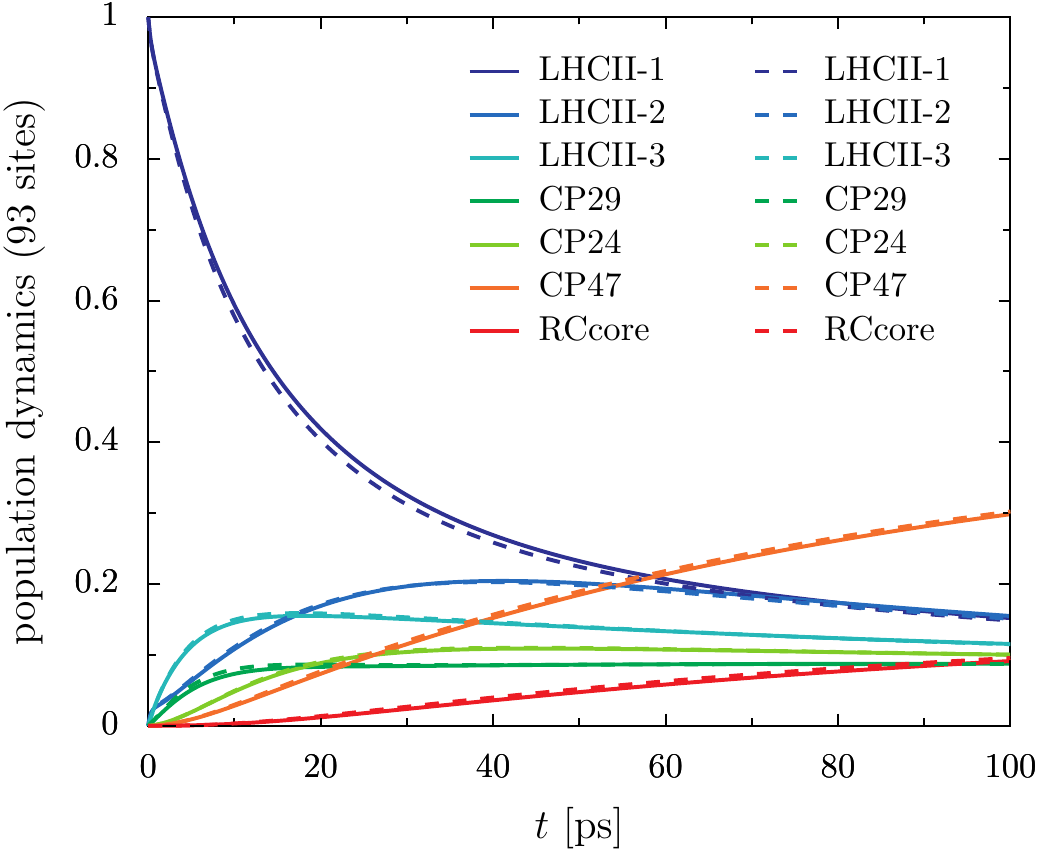}
\end{center}
\caption{\label{fig:Plot_Module_TestConvergence}\small HEOM results for the aggregated populations at 277~K for different hierarchy truncation levels $N_{\rm max}=2$ (solid lines) and $N_{\rm max}=3$ (dashed lines).}
\end{figure}

\end{suppinfo}

\end{document}